\documentclass
[twocolumn,aps,prc,amsmath,showpacs,amssymb,floatfix]
{revtex4}
\usepackage{amssymb}

\usepackage{CJK}                     
\usepackage[dvips]{graphicx}
\usepackage{bm}                      
\usepackage{mathptmx}                
\usepackage{dcolumn}                 
\usepackage{array}
\begin{document}

\title{Extended quark mean-field model for neutron stars}
\author{J. N. Hu$^{2}$, A. Li$^{1,3}$\footnote{liang@xmu.edu.cn}, H. Toki$^{4}$, W. Zuo$^{3,5}$}
\affiliation{$^1$ Department of Astronomy and Institute of Theoretical Physics
and Astrophysics, Xiamen University, Xiamen, Fujian 361005, China\\
$^2$ State Key Laboratory of Nuclear Physics and
Technology, School of Physics, Peking University, Beijing 100871,
China\\
$^3$ State Key Laboratory of Theoretical Physics, Institute of
Theoretical Physics, Chinese Academy of Sciences, Beijing 100190,
China\\
$^4$ Research Center for Nuclear Physics (RCNP), Osaka University,
Ibaraki, Osaka 567-0047, Japan\\
$^5$ Institute of Modern Physics, Chinese Academy of Sciences,
Lanzhou 730000, China\\
}
\date{\today}

\begin{abstract}

We extend the quark mean-field (QMF) model to strangeness freedom to study the properties of hyperons ($\Lambda,\Sigma,\Xi$) in infinite baryon matter and neutron star properties. The baryon-scalar meson couplings in the QMF model are determined self-consistently from the quark level, where the quark confinement is taken into account in terms of a scalar-vector harmonic oscillator potential. The strength of such confinement potential for $u,d$ quarks is constrained by the properties of finite nuclei, while the one for $s$ quark is limited by the properties of nuclei with a $\Lambda$ hyperon. These two strengths are not same, which represents the SU(3) symmetry breaking effectively in the QMF model. Also, we use an enhanced $\Sigma$ coupling with the vector meson, and both $\Sigma$ and $\Xi$ hyperon potentials can be properly described in the model. The effects of the SU(3) symmetry breaking on the neutron star structures are then studied. We find that the SU(3) breaking shifts earlier the hyperon onset density and makes hyperons more abundant in the star, in comparisons with the results of the SU(3) symmetry case. However, it does not affect much the star's maximum mass. The maximum masses are found to be $1.62~M_{\odot}$ with hyperons and $1.88~M_{\odot}$ without hyperons. The present neutron star model is shown to have limitations on explaining the recently measured heavy pulsar.
\end{abstract}


\maketitle
\section{Introduction}

Hyperon-meson couplings and their repulsive/attactive natures, are
crucial for hypernuclei physics and neutron star (NS) properties in
relativistic effective field theories, such as the relativistic mean
field (RMF) model~\cite{Sch92,Sch00,Vid00,Sch02,shen02c,shen06,Li07,Long12,Wei12,Tsu13,Col13,Pro13,Miy13},
the quark-meson-coupling (QMC) model~\cite{Fle90,Sai94,Gui96,Sai97,Pan97,Pan99,Pan03,Pan04,shen06c,Pan12}, and the quark mean field (QMF) model~\cite{Tok98,shen00,shen02,Wang01,Wang02,Wang03,Wang04,Wang04c,Wang05,Wang05c}.

For the case of hypernuclei, they essentially determine whether there is a possibility of the production of the relevant hypernuclei in the laboratory. For example, $\Lambda$-nucleus~\cite{L} and $\Lambda\Lambda$ interactions~\cite{LL} are long known as attractive ones, while an opposite sign is indicated for the $\Sigma$-nucleus interaction (see e.g. \cite{Sch00,Li07}). Recently an attractive nature has been suggested for the $\Xi$-nucleus interaction~\cite{E885,Hiy08}. For example, the BNL-E885 collaboration measured the missing mass spectra for the $^{12}$C$(K^-, K^+)$X reaction~\cite{E885}, and reasonable agreement between this data and theory is realized by assuming a $\Xi$-nucleus Wood-Saxon potential with a depth of $-14$ MeV. Within the realistic Nijmegen ESC08 baryon-baryon interaction models~\cite{Rij10}, $\Xi$-nucleus for low densities are also found to be attractive enough to produce $\Xi$ hypernuclear states in finite systems~\cite{Rij13,Yam13}. Nowadays, $\Xi$ hypernulcei have been planned for several radiation active beam factories around the world (for example, in the Japan Proton Accelerator Research Complex (J-PARC)). They are very promising objects that will contribute significantly to understanding nuclear structure and interactions in $S=-2$ systems, giving us more insight into the general understanding of the baryon-baryon interaction, as many successfully-produced $\Lambda$ hypernuclei have done.

Therefore, any effective many-body theories should respect those hypernuclei data before proceeding other sophisticated studies. The adopted hyperon-meson couplings need at least reproducing unambiguous hypernuclear data, for example, the single $\Lambda$ potential well depth in symmetric nuclear matter, $U^{(N)}_{\Lambda} \sim -30$ MeV~\cite{L}. Especially, the usually employed flavor SU(3) symmetry, as a way to determine hyperon couplings from the corresponding nucleon coupling, may have to be loosed~\cite{Tsu13},
since the construction of realistic hyperon interactions has already been performed based on a broken flavor SU(3) symmetry~\cite{Rij10}.

Furthermore, one can constrain more microscopically the hyperon-scalar couplings consistently from the quark level. Regarding to this issue, the QMC model and the QMF model can serve equivalently well in a different manner. These two models have the same root from the Guichon model proposed in 1988~\cite{Gui88}, where the meson fields couple not with nucleons as in the RMF theory~\cite{rmf}, but directly with the quarks in nucleons, then the nucleon properties change according to the strengths of the mean fields acting on the quarks, allowing us to study properties of nuclear many body systems directly from a phenomenological model of the quark-quark confinement potential. Before doing that, a nucleon model is necessary. Two nucleon models available, namely the MIT bag model~\cite{bag} and the constituent quark model~\cite{cons}, were finally developed as the QMC model and the QMF model, respectively. For a more detailed comparison of these two models we refer to Ref.~\cite{shen00}.
Shortly speaking, the first model assumes the nucleon constitutes bare quarks in the perturbative vacuum, i.e. current quarks, with a bag constant to account for the energy difference between perturbative vacuum and the nonperturbative vacuum, while in the second one, the nucleon is described in terms of constituent quarks, which couple with mesons and gluons.

The QMC model has been generalized by Fleck et al.~\cite{Fle90} and Saito and
Thomas~\cite{Sai94}, and employed extensively in many calculations of finite nuclei and infinite nuclear matter
\cite{Gui96,Sai97,Pan97,Pan99,Pan03,Pan04,shen06c,Pan12}.
The QMF model has been applied to nuclear matter~\cite{shen00} and then to finite nuclei~\cite{shen02}. More recently, Wang et al.~\cite{Wang01} included the chiral symmetry in the QMF model, and it is then called by authors as 'the chiral SU(3) QMF model', or 'a QMC model based on SU(3)$_L$ $\times$ SU(3)$_R$ symmetry'.

In this chiral SU(3) QMF model~\cite{Wang01}, an effective chiral Lagrangian was introduced with an explicit symmetry$-$breaking term for reproducing the reasonable hyperon potentials in hadronic matter. They use two parameters $h_1$ and $h_2$ to achieve an overall good agreement of all the hyperon potentials for the four types of quark confinement potentials.
In Ref.~\cite{Wang04}, they further introduced an linear definition of the effective baryon mass to postpone the critical density of a zero effective baryon mass (i.e., achieve a slower decreasing of the mass at high density), than the usual square root ansatz.
This linear definition of the effective mass was applied to a NS study in Ref.~\cite{Wang05c}, together with a scalar confining potential. In their calculation, the values of a single hyperon in nuclear matter are obtained as $U^{(N)}_{\Lambda} =U^{(N)}_{\Sigma} = -28$ MeV and $U^{(N)}_{\Xi}$ = $8$ MeV.
They finally got a maximum mass of $1.45~M_{\odot}$ ($1.7~M_{\odot}$) with (without) hyperons in the star's dense core.

In the present work, based on our previous studies ~\cite{shen00,shen02},
an extended-QMF (EQMF) model is formulated to the study of the properties of hyperons ($\Lambda,\Sigma,\Xi$) in infinite nuclear matter and NSs. Special efforts are devoted to introduce
effectively the SU(3) symmetry breaking in a different way of Ref.~\cite{Wang01}. That is, we do not include explicit symmetry$-$breaking term~\cite{Wang01} in the effective Lagrangian. Instead, we assume a different confining strength for the $s$ quark with the $u,d$ quarks in the corresponding Dirac equations (under the influence of the meson mean fields).
And the confining strength of $u,d$ quark is constrained from finite nuclei properties, and that of the $s$ quark by the well-established empirical value of $U^{(N)}_{\Lambda} \sim -30$ MeV.
The presently expected single $\Sigma$ potential of $U^{(N)}_{\Lambda} \sim 30$ MeV~\cite{Sch00} is then used to determine the $\Sigma$ coupling with vector $\omega$ meson. Namely, slightly larger $\Sigma-\omega$ coupling has been taken, as compared to $\Lambda-\omega$ coupling, to simulate the additional repulsion on the $\Sigma$-nucleon channel.

We use a scalar-vector type of harmonic oscillator potential for the confinement, instead of a scalar one used in Ref.~\cite{Wang05c},
since a denser matter can be achieved before the effective mass drop to zero (shown in Ref.~\cite{shen00}), which serves our purpose of the study of NSs with hyperon cores. Also, based on those fairly developed model calculations, we also try to emphasize some general features of relativistic models widely used in the literatures, and contribute a more comprehensive understanding of effective many-body theories. Moreover, since we do connect the theoretical NS maximum mass with the underlying quark-quark confining potentials, an analysis of their dependence is feasible, and also the theoretical implications to the recent NS mass measurements.

The paper is organized as follows. In Sec. II, we demonstrate how
the EQMF model is obtained incorporating all eight octet baryons,
including a differently-modeled $s$ quark potential strength,
the consistent determination of baryon-scalar coupling from the
quark level, and the consequential description of NS properties; The numerical results and discussions are given in Sec. III. Finally, Sec. IV contains the
main conclusions and future perspectives of this work.

\section{Formalism}

We shall begin with a possible Lagrangian~\cite{Tok98,shen00,shen02,Hu13ptep} of the quark many body system, taking into account the nonperturbative
gluon dynamics of spontaneous chiral symmetry breaking
and quark confinement. In this effective Lagrangian, we construct the interaction between baryons through the meson fields, $\sigma$, $\omega$ and $\rho$. The nucleon and meson fields are treated as mean field approximation. The inclusion of other mesons is straightforward. Then in the second step, we solve the entire baryon system by knowing the individual baryon properties due to the presence of the mean fields.

In the first step,  octet baryons ($N, \Lambda, \Sigma,
\Xi$) are described as composites of three quarks satisfying the Dirac equations with confinement potentials. The Dirac equations for constituent quarks can be written as:
\begin{eqnarray}\label{dirac}
 \left[ -i \vec \alpha \cdot \vec\nabla + \beta m_i^{*} +
 \beta \chi_c^i \right] q^i(r)=e_i^{*} q^i(r),
 \end{eqnarray}
where $i=q,s$ with the subscript $q$ denotes $u$ or $d$ quark. The
quark masses, $m_q=313\;\rm{MeV}$ and $m_s=490\;\rm{MeV}$, are
modified to $m_i^{*}=m_i+g^i_{\sigma}\sigma$ due to the presence of
the $\sigma$ mean field.
$e_i^{*}=e_i-g^i_{\omega}\omega-g^i_{\rho}\rho\tau^i_3$, with
$\sigma$, $\omega$, and $\rho$ being the mean fields at the middle
of the baryon. $e_i$ is the energy of the quark under the influence
of the $\sigma$, $\omega$, and $\rho$ mean fields. The confinement
potential is chosen to be a scalar-vector one as
$\chi^i_c=\frac{1}{2}k^ir^2 (1+\gamma^0)/2$. For the potential
strength, a previous study~\cite{shen02} of $\Lambda$ hypernuclei chose
$k^q=k^s=700 \;\rm{MeV/fm^2}$, applying the SU(3) symmetry. We here
respect the difference between $u,s$ quarks and $s$ quark, and
adjust $k^s$ to reproduce properly hypernuclei experimental data.
Then we generate the mass difference among baryons by taking into
account the spin correlations:
$E_B^{*}=\sum_ie_i^{*}+E^B_{\rm{spin}}$, where $B = N,\Lambda,
\Sigma, \Xi$. The spin correlations are fixed by fitting the baryon
masses in free space, namely $M_N=939\;\rm{MeV}$,
$M_{\Lambda}=1116\;\rm{MeV}$,~$M_{\Sigma}=1192\;\rm{MeV}$,~$M_{\Xi}=1318\;\rm{MeV}$.
We get $E^N_{\rm{spin}} = 795\;\rm{MeV}$,$~E^{\Lambda}_{\rm{spin}} = 821\;\rm{MeV}$,
$~E^{\Sigma}_{\rm{spin}} = 759\;\rm{MeV}$, and $ E^{\Xi}_{\rm{spin}} = 825\;\rm{MeV}$ at $k^s=1100$ MeV/fm$^2$, where the single $\Lambda$ potential, $U^{(N)}_{\Lambda}\sim -30$ MeV. In
addition, the spurious center of mass motion is removed in the
usual square root method as $M_B^{*}=\sqrt{E_B^{*2}-\langle p_{\rm{c.m.}}^2
\rangle }$.

\begin{figure}
\centering
\includegraphics[width=8.5cm]{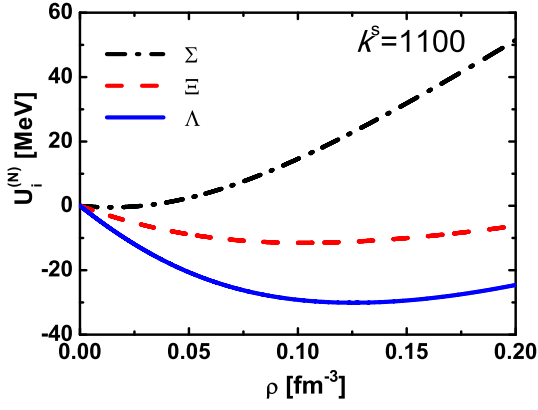}
\caption{(Color online) Single hyperon potentials $U^{(N)}_i$ as a function
of density.}\label{fig1}
\end{figure}

By solving above Dirac equations, we work out the change of the
baryon mass $M_B^*$ as a function of the quark mass correction
$\delta m_q=m_q-m_q^*$, which is used as input in the next step of
the study of nuclear many body system, that is, infinite strange
nuclear matter. Baryons inside the matter interact through exchange
of $\sigma, \omega, \rho$ mesons, and the corresponding Lagrangian
can be written as
\begin{eqnarray}\label{qmf}
 {\cal L}_{QMF}= &&\sum_B\bar \psi_B \left[ i \gamma _ \mu \partial ^\mu -
M_B^* - g_{\omega B} \omega \gamma^0 -g_{\rho B} \rho
\tau_{3B}\gamma^0\right] \psi_B  \nonumber \\
\nonumber
 & &
 -\frac{1}{2} (\bigtriangledown\sigma)^2
-\frac{1}{2} m_\sigma^2\sigma^2 -\frac{1}{4} g_3\sigma^4
 \\
\nonumber
 & &+\frac{1}{2} (\bigtriangledown\omega)^2 +\frac{1}{2}
m_\omega^2\omega^2 +\frac{1}{4} c_3\omega^4  \\
 & &+\frac{1}{2}
(\bigtriangledown\rho)^2 +\frac{1}{2} m_\rho^2\rho^2
 \end{eqnarray}
where $\psi_B$ are the Dirac spinors of baryon $B$ and $\tau_{3B}$
is the corresponding isospin projection. $m_{\sigma}$, $m_{\omega}$,
and $m_{\rho}$ are the meson masses. The mean field approximation
has been adopted for the exchanged $\sigma$, $\omega$, and $\rho$
mesons, while the mean field values of these mesons are denoted by
$\sigma$, $\omega$, and $\rho$, respectively. The contribution of
$\sigma$ meson is contained in $M_B^*$, and $\omega$ and $\rho$
mesons couple to baryons with the following coupling constants:
\begin{eqnarray}
&&g_{\omega N}=3g_\omega^q,\quad g_{\omega \Lambda}=cg_{\omega\Sigma}=2g_\omega^q,\quad
g_{\omega \Xi}=g_\omega^q \\
 & & g_{\rho N}=g_\rho^q,\quad g_{\rho \Lambda}=0,\quad g_{\rho \Sigma}=2g_\rho^q,\quad
 g_{\rho \Xi}=g_\rho^q
\end{eqnarray}
The basic parameters are the quark-meson couplings ($g^q_\sigma$,
$g_\omega^q$, and $g_\rho^q$), the nonlinear self-coupling constants
($g_3$ and $c_3$), and the mass of the $\sigma$ meson ($m_\sigma$),
which are given in Ref.~\cite{shen00} with $k^q$ = 700 MeV/fm$^{-2}$.
The saturation properties of nuclear matter with such a parameter set are listed in Table I.
As done in our previous work~\cite{Hu13ptep}, a factor $c$ is introduced before $g_{\omega\Sigma}$ for a large $\Sigma-\omega$ coupling. From reproducing the presently expected single $\Sigma$ potential $U_{\Sigma}^{(N)}$ = 30 MeV at nuclear saturation density~\cite{Sch00}, we choose $c = 0.785$ ($0.772$) for $k^s$ = 700 MeV/fm$^{-2}$ (1100 MeV/fm$^{-2}$). When $c = 1$ it goes back to the quark counting rule usually employed.

\begin{table}[t!]
\begin{center}
\caption{Saturation properties of nuclear matter used to determined
the free parameters ($g^q_\sigma$, $g^q_\omega$, $g^q_\rho$, $g_3$,
$c_3$, $m_{\sigma}$) in the present model. The saturation
density and the energy per particle are denoted by $\rho_0$ and
$E/A$, and the incompressibility by $K$, the effective mass by
$M_n^*$, the symmetry energy by $a_{sym}$.} \vspace{5pt}
\begin{tabular}{ccccc}
\hline\hline
~~$\rho_0$~~&~~$E/A$~~&~~$K$~~&~~$M_n^*/M_n$~~&~~$a_{sym}$~~
\\
~~(fm$^{-3}$)~~&~~(MeV)~~&~~(MeV)~~&~~~~~~&~~(MeV)~~
\\ \hline
 0.145 & -16.3 & 280 & 0.63 & 35
\\ \hline\hline
\end{tabular}\label{nucl}
\end{center}
\end{table}

For infinite matter, introducing the mean field approximation, we
can write the equations of motion from the Lagrangian given in
Eq.~(\ref{qmf}) as
\begin{eqnarray}
\label{eq:m1} && m_\sigma^2\sigma+g_3 \sigma^3
= \sum_B \frac{\partial M_B^*}{\partial \sigma} \frac{%
2J_B + 1}{2}\rho^s_B, \\
\label{eq:m2} & & m_\omega^2\omega+c_3 \omega^3 = \sum_B g_{\omega B}
\frac{%
2J_B + 1}{2}\rho_B, \\
\label{eq:m3}& & m_\rho^2\rho = \sum_B g_{\rho B} I_{3B}\frac{%
2J_B + 1}{2}\rho_B.
\end{eqnarray}
where $J_B$ and $I_{3B}$ denote the spin and the isospin projection
of baryon $B$. And the the baryon scalar density $\rho_B^s$ is
defined as \begin{eqnarray}
\label{eq:s}\rho^s_B=\frac{1}{\pi^2}\int\limits_0^{k_{B}} dk\,k^2\frac{M^*_B}{\sqrt {%
M_B^{*2}+k_B^2}}
\end{eqnarray}
with $k_B$ the Fermi momentum of the baryon species $B$. The total
baryon density is calculated as
$\rho=\rho_{N}+\rho_{\Lambda}+\rho_{\Sigma}+\rho_{\Xi}$.

\begin{figure}
\centering
\includegraphics[width=8.5cm]{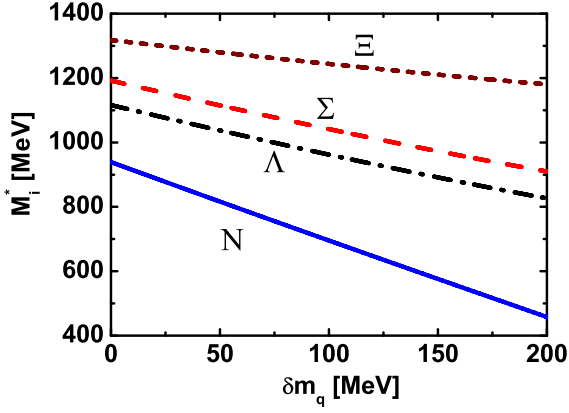}
\caption{(Color online) Effective baryon mass $M_B^*$ as a function
of the quark mass correction $\delta m_q=m_q-m_q^*=-g^i_{\sigma}\sigma$.}\label{fig2}
\end{figure}

To add leptons $L_{l}=\sum_{L=e,\mu}\overline{\psi}_{L}(i\gamma
^{\mu }\partial_{\mu }-m_{L})\psi_{L}$ to the above Lagrangian of
hadronic matter (Eq.~(\ref{qmf})), the charge neutrality requires:
\begin{equation}
\label{eq:c}
\rho_p+\rho_{\Sigma^+}=\rho_e+\rho_{\mu}+\rho_{\Sigma^-}+\rho_{\Xi^-}
\end{equation}
and equilibrium under the weak process ($B_1$ and $B_2$ denote
baryons)
\begin{equation} B_1\rightarrow B_2+L
\hspace{5mm}B_2+L\rightarrow B_1 \nonumber
\end{equation}
leads to the following relations among the involved chemical
potentials:
\begin{eqnarray} \label{eq:mu}
&&\mu_p=\mu_{\Sigma^+}=\mu_n-\mu_e,
\hspace{1mm}\mu_{\Lambda}=\mu_{\Sigma^0}=\mu_{\Xi^0}=\mu_n  \\
\nonumber &&\mu_{\Sigma^-}=\mu_{\Xi^-}=\mu_{n}+\mu_{e}, \hspace{1mm}
\mu_{\mu}=\mu_{e}.
\end{eqnarray}
where $\mu_i$ is the chemical potential of species $i$.

We solve the coupled Eqs.~(\ref{eq:m1}), (\ref{eq:m2}), (\ref{eq:m3}), (\ref{eq:c}), and (\ref{eq:mu}) at a given baryon density $\rho$, with the effective masses $M_B^{*}$ obtained at the quark level. The equation of state (EoS) of the system can be calculated in the standard way. The stable configurations of a NS then can be obtained from the well known hydrostatic equilibrium equations of Tolman, Oppenheimer and Volkoff~\cite{Tolman:1939,Oppenheimer:1939,Shapiro:1983} for the pressure $P$ and the enclosed mass $m$
\begin{equation}
 \frac{dP(r)}{dr}=-\frac{Gm(r)\mathcal{E}(r)}{r^{2}}
 \frac{\Big[1+\frac{P(r)}{\mathcal{E}(r)}\Big]
 \Big[1+\frac{4\pi r^{3}P(r)}{m(r)}\Big]}
 {1-\frac{2Gm(r)}{r}},
    \label{tov1:eps}
\end{equation}
\begin{equation}
\frac{dm(r)}{dr}=4\pi r^{2}\mathcal{E}(r),
    \label{tov2:eps}
\end{equation}
once the EoS $P(\mathcal E)$ is specified, being $\mathcal E$ the
total energy density ($G$ is the gravitational constant). For a
chosen central value of the energy density, the numerical
integration of Eqs.(\ref{tov1:eps}, \ref{tov2:eps}) provides the
mass-radius relation. For the description of the NS's crust, we have
joined the hadronic EoSs above described with the ones by Negele and
Vautherin~\cite{negele:1973} in the medium-density regime
($0.001~$fm$^{-3}<\rho<0.08~$fm$^{-3}$), and the ones by
Feynman-Metropolis-Teller~\cite{feynman:1949} and
Baym-Pethick-Sutherland~\cite{baym:1971} for the outer crust
($\rho<0.001$~fm$^{-3}$).

\section{Results}

\begin{figure}
\centering
\includegraphics[width=8.5cm]{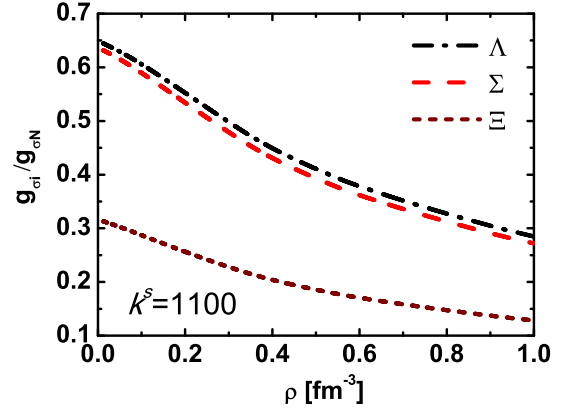}
\caption{(Color online) $g_{\sigma B}/g_{\sigma N}$ as a function of
the baryon density $\rho$ for beta equilibrium matter, with $g_{\sigma B}$ defined by $\partial M_B^*/\partial \sigma$ in the present QMF model.} \label{fig3}
\end{figure}

The potential strength of strange quark, $k^s$ must be equal to the strength of $u,d$ quark, $k^q$, if the SU(3) symmetry is considered. However, the SU(3) symmetry is not strict conserved in nuclear physics, e.g. there is a mass difference between $\Lambda$ and $\Sigma$ hyperon. Therefore, the strange potential strength, $k^s$ will differ from the $u,d$ quark case to take the effect of SU(3) symmetry breaking. $k^q$ in the QMF model is determined by the ground state properties of finite nuclei. Similarly, the magnitude of $k^s$ can be extracted from the properties of hypernulcei, such as $\Lambda$ hypernuclei, which is well-known in the strangeness physics. Its single particle potential, $U^{(N)}_\Lambda$, is around $-30$ MeV, at nuclear saturation density. With such a constraint, we can choose the strange potential strength in the QMF model as, $k^s=1100$ MeV/fm$^2$, which can generate the single $\Lambda$ potential as $U^{(N)}_\Lambda=-29.64$ MeV at the saturation density, $\rho=0.145$ fm$^{-3}$. While this value is only $U^{(N)}_\Lambda=-25$ MeV when an equal value of $k^q = k^s = 700$ MeV/fm$^2$ is chosen as done in the previous study~\cite{shen02}. Meanwhile we have checked that a reasonable description of baryon radii around $0.6$ fm is fulfilled.

In Fig.~\ref{fig1}, the single hyperon ($\Lambda,\Sigma,\Xi$) potentials as a function of density are plotted with $k^s=1100$ MeV/fm$^2$.
With density increasing, the single hyperon potentials are reduced as same as the nucleon case, which is caused by the repulsive effect stronger at high density. Furthermore, the $\Xi$ hypernuclei is possible to exist from our attractive $\Xi$ potential, although such bound state is a little bit weak about $10$ MeV, consistent with the experiments~\cite{E885}. However, we notice that in the study of the SU(3) QMF model~\cite{Wang05c}, a repulsive $\Xi$ potential is obtained with $U^{(N)}_\Xi=8$ MeV.
As for the $\Sigma$ potential, it is always repulsive as caused by the use of slightly larger $\omega$ coupling strength. This is consistent with the experimental facts that no middle and heavier mass $\Sigma$-hypernuclei have been found. The different hyperon potentials will manifest themselves in the relevant fractions of the stellar matter, as shown later.

\begin{figure}
\centering
\includegraphics[width=0.5\textwidth]{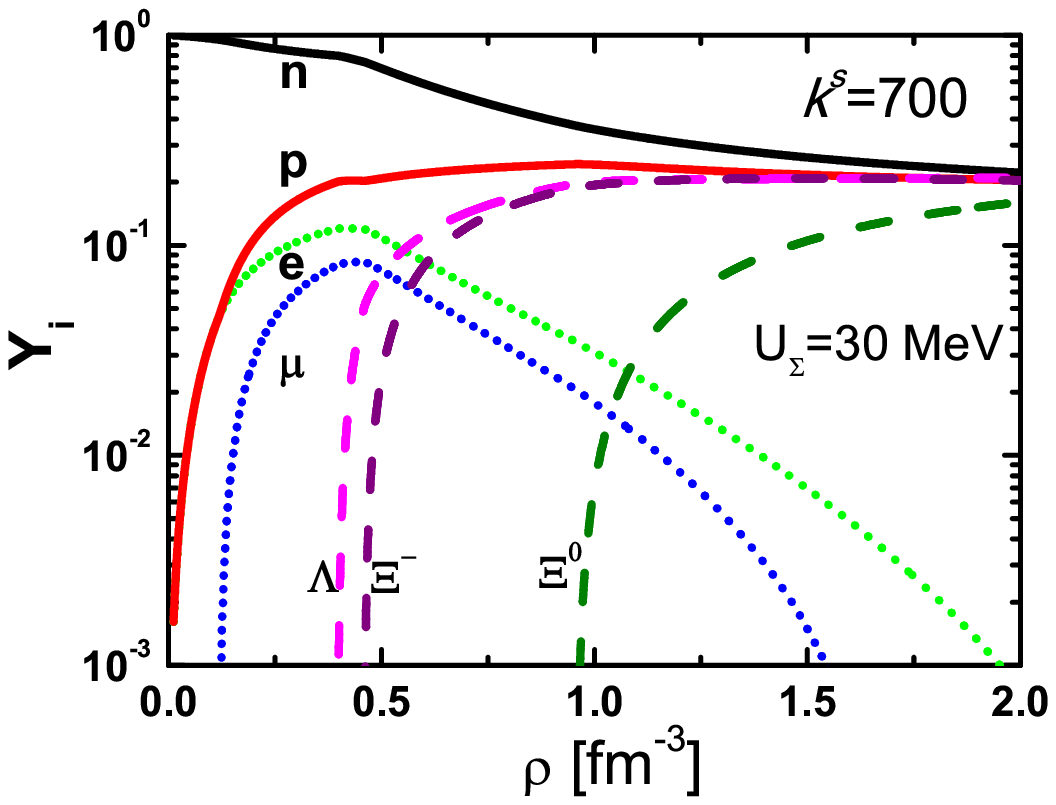}
\includegraphics[width=0.5\textwidth]{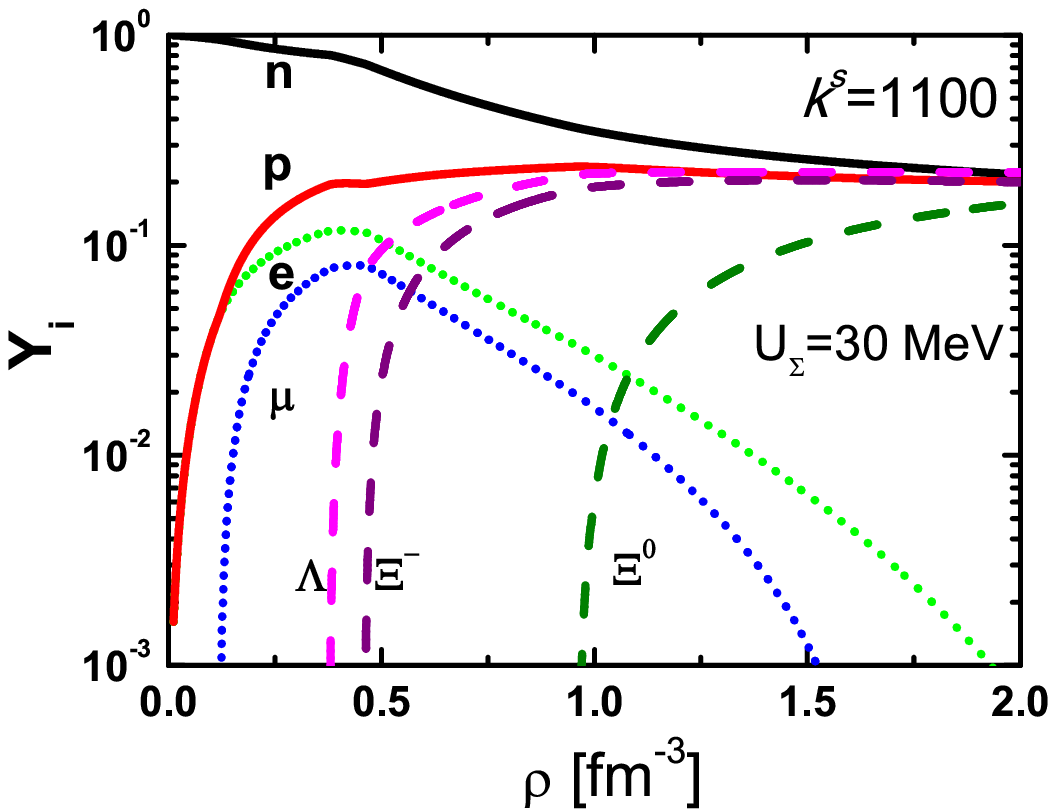}
\caption{(Color online) Fractions of leptons and baryons in NS matter are shown as function of total baryon density, for both (upper panel) $k^s=700$ MeV/fm$^2$ and (lower panel) $k^s=1100$ MeV/fm$^2$ cases.}\label{fig4}
\end{figure}

Once the strange potential strength, $k^s$, is known, we can calculate the effective baryon mass $M_B^*$ by solve the Dirac equation, namely Eq.(\ref{dirac}). The baryon masses $M_B^*$ of $\Lambda, \Sigma$ and $\Xi$ is given in Fig.~\ref{fig2} as functions of the quark mass correction. They are almost linear with the quark mass correction. Such behavior is strongly dependent on the form of quark potential as shown in Ref.~\cite{shen00}, and a linear relation is expected if a scalar-vector confining potential is employed. The hyperons in many-body system will be influenced by the surrounded hyperons and nucleons, and it is reflected in the effective hyperon masses shown here.

In the QMF model, the hadron part is dealt with the RMF theory~\cite{rmf}. The interaction between baryons in the RMF theory are provided by meson exchanges. The coupling between $\sigma$ meson and baryon can be extracted from the baryon structure in the QMF model. They are strongly dependent on the baryon effective mass $\partial M_B^*$ as defined by $g_{\sigma B}=\partial M_B^*/\partial \sigma$. The ratios of $g_{\sigma \Lambda}$, $g_{\sigma \Sigma}$, $g_{\sigma \Xi}$ to $g_{\sigma N}$ are shown in Fig.~\ref{fig3} as a function of the total baryon density $\rho$ for beta equilibrium matter. At very low density, these ratios almost satisfy the quark counting rules, approaching $2/3$ for $\Lambda, \Sigma$ and $1/3$ for $\Xi$. While with the increase of density all of them decrease steadily. This density dependent behavior shows that the effect of strange quark is weaker at high densities. Furthermore, we also notice that there is a small difference between the ratios of $\Lambda$ and $\Sigma$ indicating the SU(3) symmetry breaking.

Solving the beta equilibrium conditions in NS matter, we obtain the fraction of species, $i$, $Y_i=\rho_i/\rho$, as a function of total baryon density $\rho$ as given in Fig. \ref{fig4}. At low density region (until $\rho <0.21$ fm$^{-3}$), the proton fraction $\frac{\rho_p}{\rho_n+\rho_p}$ is well below $1/9$, which fulfill the astrophysical observations that direct URCA cooling does not occur at too low densities.

With the properly-chosen $\Lambda, \Sigma$ and $\Xi$ hyperon potentials, $\Lambda$ is the first hyperon appearing at $\rho=0.34$ fm$^{-3}$, which is lower than the one from the SU(3) symmetry calculation, 0.40 fm $^{-3}$. Namely $\Lambda$ hyperons appear earlier in the SU(3)$-$breaking case, as a result of a larger $\Lambda-$nucleon attraction. Then $\Xi^-$ hyperons appear at $\rho=0.46$ fm$^{-3}$ followed by $\Xi^0$ hyperons at $\rho=0.96$ fm$^{-3}$. These two values change not much whether we choose the SU(3)$-$breaking potential or the SU(3) symmetry one. The fractions of hyperons increase with density. Above $\rho>1.25$ fm$^{-3}$, the fractions of $\Lambda$ and $\Xi^-$ are almost the same as the fractions of proton and neutron. $\Sigma^{-}$, however, will not appear until very high density up to $2.0$ fm$^{-3}$. The appearing hyperon sequences are essentially different with the previous calculations using the quark counting rule for $\Sigma-\omega$ coupling~\cite{shen02c}, where $\Sigma^{-}$ would be the first hyperon appearing at similar density of $\Lambda$, as in also the SU(3) QMF model~\cite{Wang05c}.

\begin{figure}
\centering
\includegraphics[width=8.5cm]{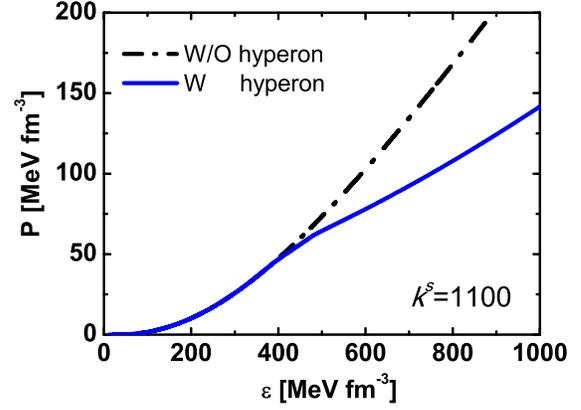}
\caption{(Color online) Pressures for beta-equilibrated matter are
shown as a function of the energy density, for both cases with or without hyperons.}\label{fig5}
\end{figure}

We also show the pressure of beta-equilibrated matter as a function of energy density in Fig.~\ref{fig5}. The solid curve represents the EoS including the hyperon and dot-dashed one is the EoS without hyperons. The EoS becomes softer after presence of the strangeness freedom.

\begin{figure}
\centering
\includegraphics[width=8.5cm]{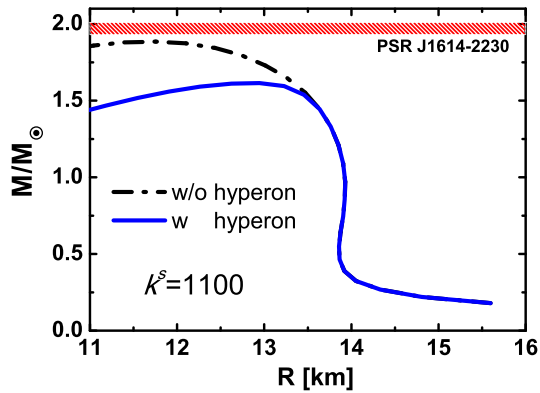}
\caption{(Color online) Gravitational masses of NSs are shown as a
function of radius, for both cases with or without hyperons in the star's core. The recently measured pulsar, PSR J1614-2230, is also indicated with a horizontal shadowed area.}\label{fig6}
\end{figure}

The NS properties are calculated by using the EoSs with/withput hyperons obtained from the EQMF theory. The NS mass-radius relations are plotted in Fig.~\ref{fig6}. It is found that the maximum mass of the NSs including hyperons is around $1.62~M_{\odot}$, while it is around $1.88~M_{\odot}$ without hyperons. Those values are larger than the corresponding results in the SU(3) QMF model mentioned before.
However, both of them could not explain the observation of $2~M_{\odot}$ NS~\cite{2times}. Our results are consistent with the conventional RMF calculations including hyperons~\cite{Sch92,Sch00,Vid00,Sch02,shen02c,Li07,Lat04}, and the microscopic studies~\cite{Sch06,Bur11,Sch11} based on developed realistic baryon-baryon interactions~\cite{Rij10}.

Since the NS maximum mass is determined by the high density region of EoS, and a stiffer EoS generates a heavier NS. It is necessary to introduce the extra repulsive mechanism in the QMF scheme, as theoretical efforts done in the RMF framework in Refs.~\cite{Tsu13,Col13,Pro13,Miy13,Wei12}. Also, in a recent work of QMC model~\cite{Pan12}, besides the usual $\sigma$, $\omega$, $\rho$ fields, a nonlinear $\omega-\rho$ term was introduced (with a new coulping parameter $\Lambda_v$) in the Lagrangian, to correct the stiff behavior of the symmetry energy at large densities. For example, the slope parameter $L$ of the symmetry energy was lowered from 93.59 MeV to 39.04 MeV for $\Lambda_v = 0.1$. As a result, they got a softer nuclear EoS at high densities (which hinders the onset of hyperons) and a harder EoS with hyperons, with the help of the reduction of the attractiveness of $\Xi$ potential $U_{\Xi}$, a $2~M_{\odot}$ NS was finally possible in the model. Similar extensions can be done in the QMF model. However, since the maximum mass of the pure NSs is as heavy as $1.88~M_{\odot}$ in the present QMF model, one would not expect the corresponding hyperon stars could be heavier than that. This demonstrates the limitations of the present neutron star model.

\begin{figure}
\centering
\includegraphics[width=8.5cm]{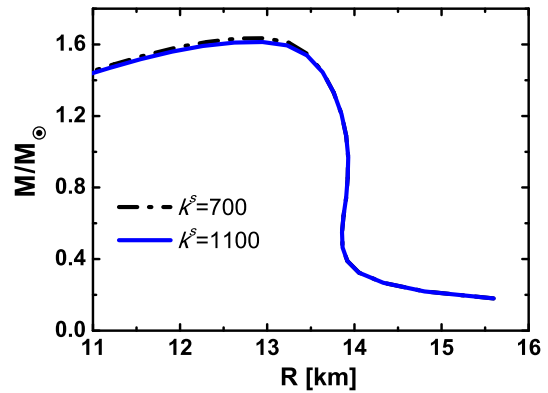}
\caption{(Color online) Gravitational masses of hyperon stars are shown as a function of radius, for both $k^s=700$ MeV/fm$^2$ and $k^s=1100$ MeV/fm$^2$ cases.}\label{fig7}
\end{figure}

To discuss the effect of SU(3) symmetry breaking on the NS structure. We also calculate the mass-radius relation of NS with $k^s=700$ MeV/fm$^{2}$. The results are plotted in Fig.~\ref{fig7}, compared with the breaking case of $k^s=1100$ MeV/fm$^{2}$. The solid curve is the mass-radius relation considering the SU(3) symmetry breaking, while the dot-dashed one is SU(3) symmetry conservation at the quark level. The maximum masses of NS is not much changed in these two cases, only slightly lowered in the symmetry breaking case resulting from more hyperon softening, as indicated in Fig.~\ref{fig4} for the compositions of the matter.

\section{Summary and future perspectives}

We extended the QMF model to study infinite hyperonic matter, which include the $\Lambda$, $\Sigma$ and $\Xi $ hyperons. The SU(3) symmetry was broken in the quark level to be consistent with the experiment data of $\Lambda$ potential at nuclear saturation density, i.e., $U^{(N)}_\Lambda\sim30$ MeV. Namely we chose different potential strengths for $u,d$ and $s$ quarks at quark mean fields.

Using such quark potential strengthes, the coupling constants between $\sigma$ meson and baryons were determined through the effective baryon masses from the Dirac equation of quarks. These coupling constants strongly depended on the density and differed from the results of the quark counting rules supported by SU(3) symmetry. We also chose a slightly larger $\omega$ coupling with $\Sigma$ hyperons, than that of $\Lambda$ hyperons,
to reproduce the presently expected single $\Sigma$ potential of $U_\Sigma ~= 30$ MeV at the nuclear saturation density.
We can then obtain also a slightly attractive $\Xi$ potential desired in the hypernuclei experiments, however missing in the previous SU(3) QMF model.

We calculate the properties of NSs with the EQMF model, and discussed the SU(3) symmetry breaking effect on the NS mass. The onset of hyperons is moved ahead using the SU(3)$-$breaking potential, and the fraction of hyperons are increased in the star. However, the maximus mass of NSs was found almost not changed, comparing with the case when we kept the SU(3) symmetry in the quark level. The maximum mass of NSs approaches $1.62~M_{\odot}$ with hyperons and $1.88~M_{\odot}$ without hyperons. These results could not explain the observation of $2~M_{\odot}$ NS observation.

In order to resolve the limitations of the model, one has to readjust all the QMF parameters from reproducing finite nuclei data, to achieve a proper new parameter set to fulfill the $2~M_{\odot}$ constrain. There is also a possibility that the phase transition to a strongly-interacting quark matter in the star's core that can support $2~M_{\odot}$ gravitational mass. These will be studied in our future works.

\begin{acknowledgments}
We would like to thank Prof. H. Shen and Prof. M. Oka for valuable discussions. The work was supported by the National Natural Science Foundation of China (Nos. 11175219, 10875151, 10905048), the Major State Basic
Research Developing Program of China (No. 2007CB815004), the
Knowledge Innovation Project (KJCX2-EW-N01) of the Chinese Academy
of Sciences, the Chinese Academy of Sciences Visiting Professorship
for Senior International Scientists (Grant No. 2009J2-26),
and the China Postdoctoral Foundation (Grant No. Science 2012M520100).
\end{acknowledgments}


\end{document}